\documentclass[twocolumn,preprintnumbers,amsmath,amssymb]{revtex4}
\usepackage{graphicx}
\usepackage{color}
\begin{document}
\newcommand{\rf}[1]{(\ref{#1})}

\newcommand{\bfomega}{ \mbox{\boldmath{$\omega$}}}

\title{Granular dynamics in compaction and
stress relaxation}

\author{ Jasna Bruji\'c$^1$, Ping Wang$^2$, Chaoming Song$^2$,
David L. Johnson$^1$,  Olivier Sindt$^1$, and Hern\'an A.
Makse$^2$ } \affiliation{ $^1$ Schlumberger Doll Research, Old
Quarry Road, Ridgefield, CT
06877\\
$^2$ Levich Institute and Physics Department, City College of New
York, New York, NY 10031 }


\begin{abstract}

Elastic and dissipative properties of granular assemblies under
uniaxial compression are studied both experimentally and by
numerical simulations. Following a novel compaction procedure at
varying oscillatory pressures, the stress response to a
step-strain reveals an exponential relaxation followed by a slow
logarithmic decay. Simulations indicate that the latter arises
from the coupling between damping and collective grain motion
predominantly through sliding. We characterize an analogous
``glass transition'' for packed grains, below which the system
shows aging in time-dependent sliding correlation functions.

\center{( Dated: Phys. Rev. Lett. {\bf 95}, 128001 (2005) )}

\end{abstract}
\maketitle


Mechanically agitated granular materials are characterized by slow
relaxation dynamics, arising from the rearrangement of the
constituent grains within the volume in which they are confined
\cite{coniglio}. This leads to a slow compaction of the system
volume, which  follows a logarithmic decay in time, as seen in
several experiments and predicted by theoretical models of
granular compaction \cite{knight,bennaim}. This property has
prompted analogies between the physics of athermal granular
materials and thermal glasses, the theory of which is better
understood at the fundamental level \cite{coniglio}.  The field of
granular matter therefore benefits from such parallels, as new
ways of investigating the system's complex properties are
discovered. An advantage of using granular materials over glasses
is that it facilitates a much easier exploration of the
microstructure through grain-grain interactions.

Logarithmic relaxation  and rate-dependent strengthening have also
been observed in {\it compressed} granular matter \cite{hartley},
although the underlying mechanism is still under much debate. The
collective rearrangement of the grains in the bulk could be
responsible for the slow relaxation, as suggested by experiments
on slowly sheared granular materials by Hartley and Behringer
\cite{hartley}. On the other hand, it is also known that aging
occurs at the  contacts between the particles \cite{aging},
which manifests itself as a logarithmic increase of the friction
coefficient between grains as a function of time. It could also be
responsible for the observed slow dynamics, as has been suggested
in experiments by Ovarlez {\it et al.} and Nasuno {\it et al.}
\cite{ovarlez,nasuno}, which show rate dependence and slow
strengthening characteristics of aging at the interparticle
contacts, respectively.

The goal of this Letter is to demonstrate the existence of slow
relaxation in the response of dense granular matter to
infinitesimal strain perturbations and to elaborate on the origin
of the dynamics. The experiments reveal a very slow stress
relaxation under a constant applied differential strain. This
behavior is well characterized by a two-step relaxation dynamics,
analogous to the slow relaxation in ``glassy'' systems
\cite{coniglio}.

We investigate this dynamics via computer simulations, which
employ various dissipative processes into the system in order to
compare their relative effects.  The results show that the main
process responsible for the logarithmic stress relaxation is the
collective particle motion and rearrangements of grains,
predominantly through sliding. We compute the fraction of sliding
particles and find that when the damping in the system exceeds a
critical value, a slow increase in the number of sliding particles
as a function of time is observed. This slow strengthening leads
to the logarithmic stress relaxation. Moreover, the system shows
the hallmark of glassy behavior: aging in the sliding correlation
function.



{\it Experimental arrangement.--} We use an INSTRON press to
measure the mechanical properties of granular assemblies confined
in a cylindrical cup and piston.  The machine was strain
controlled and enabled oscillatory, step, and ramp compressional
tests up to a maximum limit of 300kN load. 
The cylinder cup had the following dimensions: diameter $d=$5.08
cm, height $h=$7.62 cm and wall thickness $l=0.85$ cm. These
dimensions were chosen to achieve a good statistical ensemble for
nearly monodisperse glass beads of diameter (355$\pm5) \mu m$. We
perform two tests: first, we measure the strain response
(compaction) under large stress oscillations, and then we measure
the stress response under an infinitesimal step-strain
perturbation.

{\it Compaction experiments (stress-controlled).--}
Before performing the stress relaxation experiments it was
necessary to develop a method of reaching the jammed state,
ensuring that reproducible experiments can be performed.


\begin{figure}
\centering \resizebox{7.5cm}{!}{\includegraphics{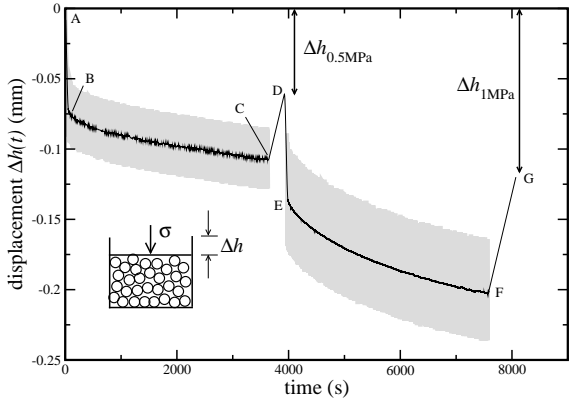}}
\caption{Compaction procedure: Sinusoidal displacement $\Delta h$
of the system height in response to an oscillatory stress around
the mean value of 0.5 MPa (from B to C) and 1 MPa (from E to F).
The solid black line is the window average displacement of the
height and shows the logarithmic behavior of the strain under
large stress amplitude according to Eq. (\protect\ref{logt}) for
the later stages of the time evolution. The shaded gray area
symbolizes the amplitude of the oscillatory strain response. The
inset shows a schematic of the experimental arrangement.}
\label{compaction}
\end{figure}

Here we introduce an alternative method to the one of Knight {\it
et al.} \cite{knight} of compaction based on oscillatory pressure
of varying amplitude to generate reversible jammed states. The
compaction procedure consists of the stages depicted in Fig.
\ref{compaction}, where we show the displacement of the piston
$\Delta h$ during the application of the stress $\sigma$. The
material is first compressed with a constant, slow velocity of 0.1
mm/min to a target stress of  0.5 MPa (from A to B in Fig.
\ref{compaction}, which shows only the strain response),
then oscillated between zero and double the mean value with a
frequency of 1Hz (from B to C, the amplitudes of oscillation are
always taken as equal to the mean value). Each compaction
procedure consists of 3,600 cycles (until C), after which the
material is again slowly released to its uncompressed state (from
Point C at $\sigma=0.5$ MPa to D at $\sigma=0$). The final height
of the material after the compaction cycle is calculated at Point
D, $\Delta h_{\mbox{\scriptsize 0.5MPa}}$, and used to obtain the
volume fraction at the given stress amplitude.  The sequence is
then repeated for increasing values of the mean stress at 1 MPa
($D\rightarrow E \rightarrow F \rightarrow G$) and the
correspondingly larger amplitudes of oscillation lead to a new
height and packing density  $\Delta h_{\mbox{\scriptsize 1MPa}}$
for the new oscillatory stress.

The dynamics of compaction is well described by a logarithmic
dependence of $\Delta h(t)$ for each compaction cycle valid for
long times:

\begin{equation}
\Delta h(t) \sim - \ln (t). \label{logt}
\end{equation}
At short times, it crosses over to a faster decay. This implies
that the density behaves as $\sim 1/\ln(t)$ at long times in
agreement with previous experiments \cite{knight}. This is shown
in Fig. \ref{compaction} as the solid line from B to C and from E
to F.
The relaxation is so slow that one could argue that the final
steady-state density has not been achieved on the timescale of the
experiment.

\begin{figure}
\centering \resizebox{7.5cm}{!}{\includegraphics{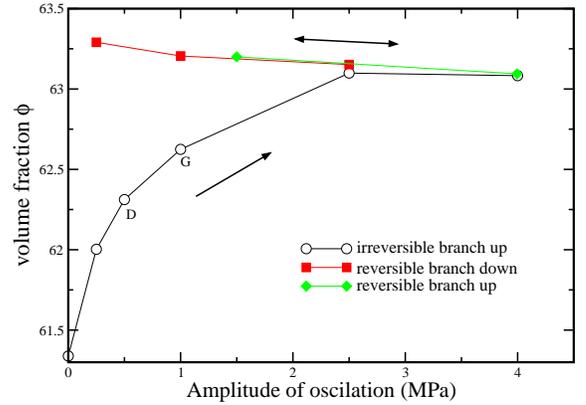}}
\caption{Volume fractions obtained by oscillatory compression
experiments at varying oscillatory stress amplitudes. The final
heights at D and G in Fig. \protect\ref{compaction} are used to
calculate the volume fractions shown in this figure.}
\label{density}
\end{figure}

The volume fraction after each compaction cycle is plotted against
the amplitude of oscillatory  stress in Fig. \ref{density}. The
protocol involves a stepwise ramp of compaction cycles from zero
stress amplitude to 4 MPa, then back to zero and back again to  4
MPa. The figure shows an irreversible branch as the mean stress is
first increased, until a plateau in the volume fraction is
reached. As the stress is then reduced the system enters into the
reversible branch of compaction. By ramping up and down in
oscillatory stress we find that the system reproduces the same
curve. These are the reversible jammed states that we use as the
reference states in the stress relaxation experiments.

{\it Stress Relaxation Experiments (strain-controlled).--} It is
important to distinguish between processes related to large scale
deformations in granular compaction studied above and
infinitesimal perturbations related to supporting the stress once
the volumetric conditions have been satisfied.  In the latter, an
application of an external stress will result in dissipation
mechanisms quite different to the compaction process.
Therefore, we next probe the mechanisms of energy dissipation of
the fully compactified system by performing infinitesimal step
compression experiments and observing the resulting response in
the stress.

We perform uniaxial compression tests with the plate-cup
configuration used above. We contract the system by applying a
step strain $\Delta \epsilon$ in the range $(1-3) \times 10^{-3}$
to the glass bead sample at a given confining pressure. Meanwhile,
we monitor the temporal evolution of the differential stress,
$\Delta \sigma(t)$. It is defined as the difference between the
stress at time $t$ measured after the strain is applied and the
stress before the perturbation is applied. It follows that
$\Delta\sigma(0)$ is the change in stress between the baseline
(stress before perturbation) and just after the perturbation is
applied \cite{Furthermore}.


\begin{figure}
\centering \resizebox{7.5cm}{!}{\includegraphics{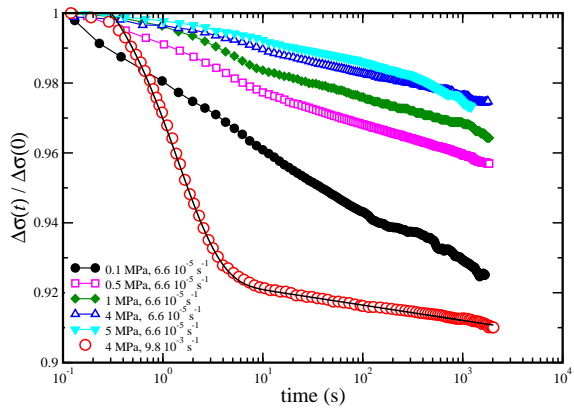}}
\caption{Experiments: stress relaxations at different confining
pressures. Results are shown for two strain rates. The fastest one
reveals the two-step relaxation. }
\label{relaxation1}
\end{figure}

Figure \ref{relaxation1} shows the resulting relaxation at
different confining pressures from 0.1 MPa to 5 MPa. At the
slowest strain rate of $\Delta \dot{\epsilon} = 6.6 \times
10^{-5}\:$s$^{-1}$ we find a relaxation which is logarithmic in
time, at long times. When we repeat the experiment at the maximum
speed allowed by the press, $\Delta  \dot{\epsilon} = 9.8 \times
10^{-3}\:$s$^{-1}$, we are able to observe the ``instantaneous"
stress response in addition to the subsequent relaxation. In this
case we find a two-step relaxation, which is well approximated
with the following equation, plotted in Fig. \ref{relaxation1} for
the 4 MPa stress relaxation:

\begin{equation}
\Delta \sigma(t) / \Delta \sigma(0) = A + B e^{-t/\tau_1}- C
\ln(t), \label{sigma}
\end{equation}
where $\tau_1=1.4$ s is the fast relaxation time and $A=0.9$,
$B=0.09$ are  constants, and $C=2\times 10^{-3}$ sets the rate of
the slow relaxation. Since the slow strain rate corresponds to a
straining time of $T=21.6$ s and the fast rate to $T=0.18$ s for
the case of the 4 MPa sample, the fast relaxation is only observed
in the latter experiment, as the relaxation time $\tau_1$ is
slower than the application of the strain.


We argue, by analogy with glassy dynamics, that the fast
relaxation is a single particle relaxation mechanism whereas the
slow is representative of a collective rearrangement of many
grains via sliding and their ensuing ``aging'' properties.  We
provide supporting evidence for this claim by means of molecular
dynamic simulations.


{\it Computer simulations.--} In order to decipher the main
microscopic mechanism, we perform a numerical study based on
Molecular Dynamics of elasto-frictional spherical particles.

Interparticle forces are computed using the principles of contact
mechanics and considers: normal Hertz forces, $F_{n}$, tangential
Mindlin forces, $F_{t}$, and dry Coulomb friction: $F_{t}\leq \mu
F_{n}$, with $\mu$ the friction coefficient. Full details are
given in \cite{mgjs2}.



We provide two principal mechanisms of dissipation.
If the grains are touching, they exert {\it contact} damping
forces, which arise from viscoelastic dissipation between the
grains:
$f^{diss}_n = - \gamma_n \xi^{1/2} \dot \xi $ and $f^{diss}_t = -
\gamma_t \xi^{1/2} \dot s$ \cite{poschel2}. Reference values for
the damping constants $\gamma_n$ and $\gamma_t$ can be  obtained
from \cite{wolf-md}. Furthermore, the grains are immersed in a
viscous fluid, such as air or water, which causes {\it global}
damping according to the classical Rayleigh theory.
The drag of a sphere immersed in a viscous fluid
is $F^{drag} = 6 \pi \eta R \dot{\mbox{\bf x}}$, where $\eta$ is
the viscosity of the fluid (an analogous expression holds for
torque damping) \cite{wolf-md}.

The packings were equilibrated at a given  pressure of 1MPa
according to the previously established method \cite{mgjs2}. We
probe the macroscopic mechanical properties by applying an
infinitesimal step strain, and we monitor the time dependence of
the stress, thus mimicking the experiments.

\begin{figure}
\centering (a)\resizebox{8cm}{!}{\includegraphics{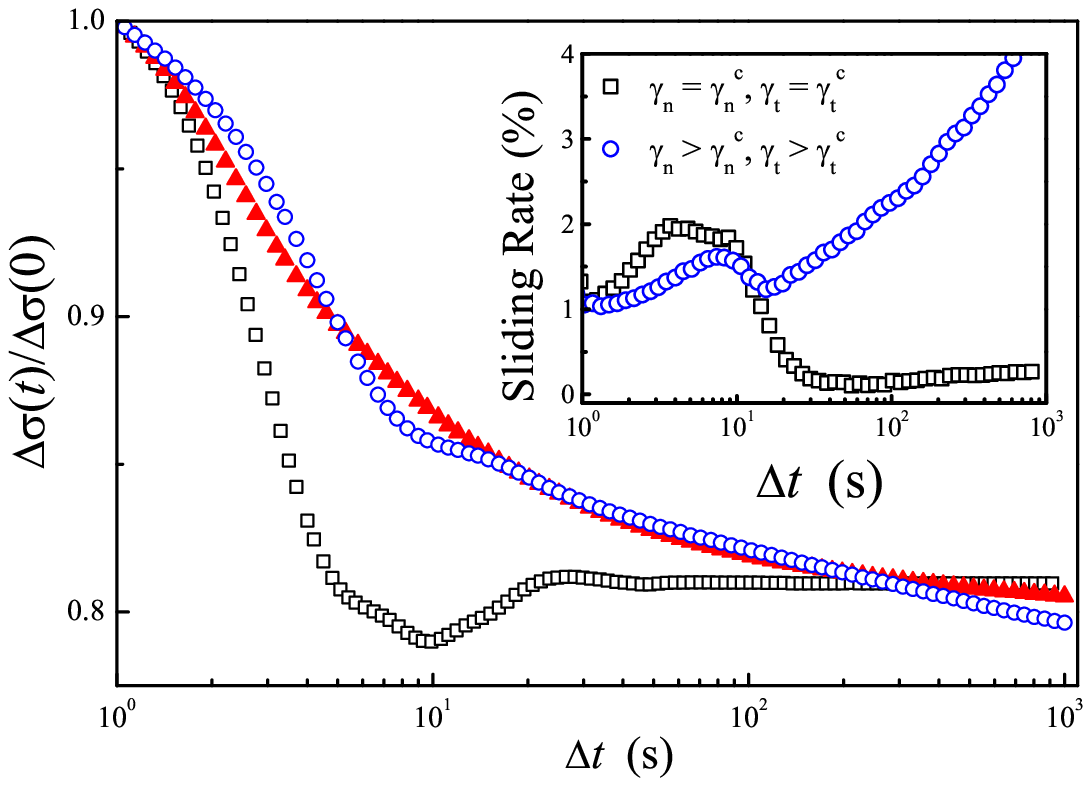}}
(b)\resizebox{8cm}{!}{\includegraphics{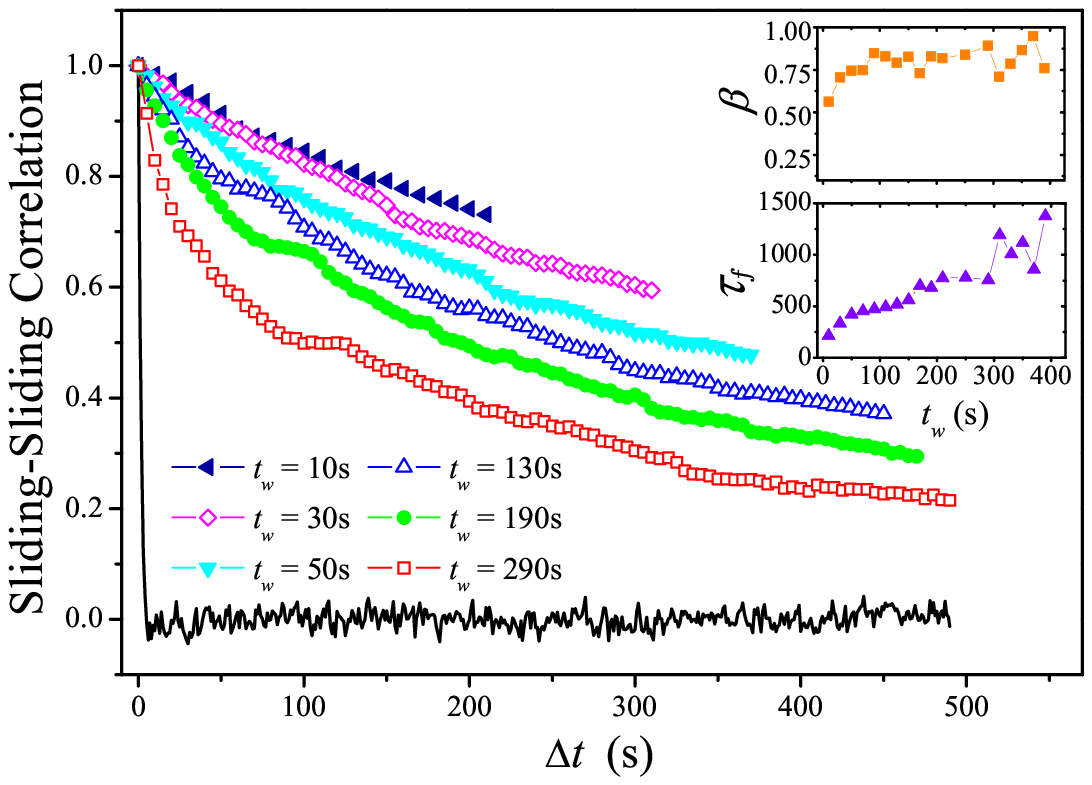}}
\caption{Simulations: (a) stress relaxation after an instantaneous
step strain at 1MPa. Red solid triangles correspond to a system
with global damping $\eta=1.7\times10^{-3}$ Pa-s $>\eta^{c}$, blue
open circles (also in the inset) to an overdamped system with
contact damping $\gamma_{n}=2\times10^{-2}\ kg/s\ m^{1/2},
\gamma_{t}=2\times10^{-3}\ kg/s\ m^{1/2}$, and black open squares
(also in the inset) to a system with critical contact damping. The
inset shows the behavior of the fraction of the particles sliding
in the aging regime and at the critical point. (b) Sliding-sliding
correlation function showing the appearance of aging (dependence
on $t_w$) above the critical damping. The black line show the
behavior below critical damping. Insets show the stretched
exponential exponent $\beta$ and the characteristic relaxation
time $\tau_f$ versus $t_w$.} \label{relaxation2}
\end{figure}

We find a critical damping (both for global and contact), above
which the slow relaxation ensues. The critical values  are
$\gamma_n^c =2\times 10^{-3}$ kg/s $m^{1/2}$, $\gamma_t^c =2\times
10^{-4}$ kg/s $m^{1/2}$, while the critical global viscosity is
$\eta^c =1.7\times 10^{-4}$ Pa-s \cite{time}. In Fig.
\ref{relaxation2}a we show typical relaxation curves obtained for
three systems with either global damping only ($\eta>\eta^c$), or
with contact damping only
($\gamma_n>\gamma_n^c,\gamma_t>\gamma_t^c$), and a third system
with critical contact damping
($\gamma_n=\gamma_n^c,\gamma_t=\gamma_t^c$), which dissipates
energy mainly via Coulomb friction. We see that the curves with
damping larger than the critical value are in good qualitative
agreement with those from the experiment, displaying slow stress
relaxation at long times, while the curve with mostly frictional
dissipation decays very fast. Comparing with real viscoelastic
constants for common glass materials \cite{wolf-md} and the
viscosity  of air or water, we find that real viscoelastic damping
is almost always above the critical values. Thus, in most
experimental situations, except perhaps in vacuum, the system will
be overdamped and the slow relaxation will be observed.


Considering that we have the full information on the motion of the
particles, we now investigate the microscopic origin of the slow
relaxation. By monitoring the small displacements of the particles
we find that the shear displacements are key to explaining the
relaxation mechanism.
The inset of Fig. \ref{relaxation2}a shows the fraction of
particles that are sliding ($F_t=\mu F_n$) in the system at a
given time (for these particles the shear displacement is
significant). If the system is underdamped ($\gamma<\gamma^c$, for
both normal and shear), this fraction decays rapidly to zero and
the stress is quickly relaxed. On the other hand, when the system
is overdamped ($\gamma>\gamma^c$) the fraction of sliding
particles increases as a function of time, analogous to the
strengthening behavior found in previous experiments
\cite{nasuno}. This slow dynamical strengthening (which may
saturate at larger times) is responsible for the
slow stress relaxation.


To further investigate the features of the overdamped state we
study the time correlation function of the sliding grains to
quantify their motion. At a given time step we construct a
state-vector $\vec s(t)$ with $M$ components ($M$ is the number of
contacts in the system); the $i$-th component being 1 or -1
according to whether the $i$-th contact is sliding or not. We
consider a time correlation function $C(t,t_w)=<\vec s(t+t_w) \vec
s(t_w)>$ with a dependence on the waiting time, $t_w$, measured
from the time when the perturbation is applied. Below the critical
damping condition we find (Fig. \ref{relaxation2}b) that the
$C(t,t_w)$ decays rapidly with no evidence of $t_w$ dependence.
With small damping forces the grains in the system have no time to
develop significant shear displacements. The number of sliding
particles is very small and the system does not display any slow
relaxation.

In contrast, above critical damping we find (Fig.
\ref{relaxation2}b) a stretched exponential decay  $C(t,t_w) \sim
\exp[-t/\tau_f(t_w)]^\beta$, where $\tau_f(t_w)$ is the
characteristic time dependent on the waiting time and
$\beta\approx 0.8$ is the critical exponent (see Insets of Fig.
\ref{relaxation2}b). Large damping sufficiently slows down the
system such that the grains have enough time to develop
substantial tangential displacements, which in turn generate shear
forces large enough to cause sliding of the grains.
Interestingly, the same effect has also been found in the
history-dependent behavior of packings near the jamming
transition, in this case, as a function of the compression rate
\cite{hepeng}.
The fact that both damping and compression can drive the system
into a glassy phase through a well-defined transition, which
exhibits aging, carries important implications.

Whereas the transition between an underdamped and overdamped state
is not surprising, it is intriguing that the latter state show
signatures of glassy behavior. Conventional glassy systems such as
polymer melts undergo the glass transition by fast cooling of the
system, while here we show that in granular athermal systems the
damping plays a similar role to temperature. A granular `glass
transition' driven by damping opens interesting unifications
between the two types of systems, the details of which we plan to
test experimentally in further work. Since damping facilitates
dissipation by friction as the timescale of grain contacts is
prolonged, we predict that a packing of frictionless droplets in
emulsions would not experience the observed slow relaxation and
aging dynamics. In fact our simulations indicate that a system
with frictionless particles, $\mu=0$, as well as a system of
infinitely rough particles,
  $\mu \to \infty$, does not display aging, thus confirming that
the glassy properties are due to the finite friction coefficient
of grains.



In summary, in this work we distinguish between stress relaxation
processes related to reaching a granular jammed equilibrium state
and the remaining infinitesimal relaxation.
Once the system is jammed at a given pressure, the stress
relaxation is characterized by a fast exponential relaxation
followed by a glassy slow logarithmic decay. The glassy phase is
characterized by the aging of the sliding correlation function and
the ensuing stretched exponential behavior, which are in turn
responsible for the slowdown of the dynamics. It's interesting to
note that the amplification of the sliding could be associated
with the initiation of an avalanche inside the system.


Acknowledgments: This work was supported by DOE Grant
DE-F602-03ER15458 and the CREST-NSF program.




\begin{thebibliography}{}

\bibitem{coniglio}
A. Coniglio, A. Fierro, H. J. Herrmann and M.  Nicodemi, (eds)
{\it Unifying Concepts in Granular Media and Glasses} (Elsevier,
Amsterdam, 2004).


\bibitem{knight}
J. B. Knight \emph{et al}., Phys. Rev. E {\bf 51}, 3957 (1995); P.
Philippe and D. Bideau, Europhys. Lett. {\bf 60}, 677 (2002); E.
R. Nowak \emph{et al}., Phys. Rev. E {\bf 57}, 1971 (1998).


\bibitem{bennaim}
P. L. Krapivsky and E. Ben-Naim, J. Chem. Phys. {\bf 100}, 6778
(1994); M. Nicodemi, A. Coniglio and H. J. Herrmann, Phys. Rev. E
{\bf 55}, 3962 (1997); T. Boutreux and P. G. de Gennes, Physica A
{\bf 224}, 59 (1997).

\bibitem{hartley}
R. R. Hartley and R. B. Behringer, Nature {\bf 421}, 928  (2003).


\bibitem{aging}
P. Berthoud, T. Baumberger, C. G'Sell and J.-M. Hiver, Phys. Rev.
B {\bf 59}, 14313 (1999).

\bibitem{ovarlez}
G. Ovarlez, E. Kolb and E. Cl\'ement, Phys. Rev. E {\bf 64},
060302(R) (2001).


\bibitem{nasuno}
S. Nasuno, A. Kudrolli, and J. P. Gollub, Phys. Rev. Lett. {\bf
79}, 949 (1997).

\bibitem{Furthermore}
Furthermore, stress is generally defined as $\Delta \sigma(t) =
C_{11}(t) \Delta \epsilon $. We thus measure a modulus which is a
function of both the compressibility ($K$) and the shear ($G$)
moduli of the material: $C_{11}(t) = K(t) + (4/3) G(t)$. Since the
compressibility is expected to have a purely elastic mechanical
response (i.e., $K(t)$= constant) \cite{mgjs2}, the relaxation can
be directly related to the shear component $G(t)$.




\bibitem{mgjs2} H. A. Makse, N. Gland, D. L. Johnson, and
L. M. Schwartz, {\it Phys. Rev. E}  {\bf 70}, 061302   (2004).

\bibitem{wolf-md} J. Sch\"afer, S. Dippel, and D. E. Wolf,
J. Phys. I (France) {\bf 6}, 5 (1996).

\bibitem{poschel2} N. V. Brilliantov, F. Spahn, J.-M. Hertzsch, and
T. P\"oschel, {\it Phys. Rev E} {\bf 53}, 5382 (1996).



\bibitem{time}
The time unit is defined such that the fast relaxation time of the
stress obtained numerically is the same as that obtained
experimentally.

\bibitem{hepeng}
H. P. Zhang and H. A. Makse, accepted in Phys. Rev. E,
cond-mat/0501370.






\end{thebibliography}
\end{document}